\documentclass[11pt]{article}

\title { \bf Complete Analytical Solutions of the Mie-type Potentials in $N$-Dimensions.}
 
\author{\bf D. Agboola\footnote{tomdavids2k6@yahoo.com}}

\date{\centering Department of Pure and Applied Mathematics,Ladoke Akintola University of Technology,Oyo State, Nigeria. \linebreak  P.M.B. 4000}

\begin{document}
\maketitle

\vspace{0.5in}
\noindent {\bf Abstract:}The exact solutions of the  $N$-dimensional Schr$\ddot{o}$dinger equation with the Mie-type potentials are obtained. The energy levels are worked out and the corresponding wave functions are obtained in terms of the Laguerre polynomial. The expectation values $\langle r^{-1}\rangle$ and $\langle r^{-2}\rangle $ and the virial theorem are also obtained in  $N$-dimensions using the Hellmann-Feynman theorem.The ladder operators are also constructed for the Mie-type potentials in $N$-dimensions and the matrix elements of some operators $r$ and $r\frac{d}{dr}$ are analytically obtained from the ladder operators.And the general results reduce to the 3-dimensional case when  $N=3$.          

\maketitle
\vspace{0.5in}
\vspace{1in}
\maketitle


\pagebreak
\noindent { \bf 1.0 Introduction}\\

 The exact bound-state solutions of the Schr$\ddot{o}$dinger equation with physically significant potentials play a major role in quantum mechanics. Over the decades, exact solutions of the multidimensional Schr$\ddot{o}$dinger equation have attracted much interest. Problems involving the N-dimensional Schr$\ddot{o}$dinger equation have been severally solved by some researchers with the special transformation of the $N$-dimensional Schr$\ddot{o}$dinger equation. For instance, Bateman investigated the relationship between the hydrogen atom and a harmonic oscillator potential in arbitrary dimensions [1]. The $N$-dimensional Kratzer-Fues potential was discussed by Oyewumi [2].Also, the $N$-dimensional Pseudoharmonic oscillator was discussed by Agboola \textsl {et al} [3]. Recently, a $D$-dimensional study of both the Hulth$\acute{e}$n [4, 5] and P$\ddot{o}$schl-Teller [6] potentials have been discussed by Agboola.   

Recently, Chen and Dong [7] found a new ring-shaped potential and obtained the exact solution of the Schr$\ddot{o}$dinger equation for the Coulomb potential plus this new ring-shaped potential which has possible applications to ring-shaped organic molecules like cyclic polyenes and benzene.Very recently, Cheng and Dai [8], proposed a new potential consisting from the modified Kratzer potential [9] plus the new proposed ring-shaped potential, and they presented the energy eigenvalues for this proposed exactly-solvable non-central potential in three dimensional Schr$\ddot{o}$dinger equation. 

The path integral solution for one-dimensional special case Mie-potential which is a perturbed
Coulombic-type potential was obtained in [34]. Moreover, the Schr$\ddot{o}$dinger equation for a system bound by a Mie-type potential was also solved by using the $1/N$ expansion method [33].
In this paper,we give a complete normalized polynomial solution for the general $N$-dimensional Schr$\ddot{o}$dinger equation for diatomic molecular systems interacting through Mie type potential, which reduces to the standard three dimensional case when the parameter $N$ is set equal to 3. Moreover,we also present some quantum-mechanical properties such as some expectation values and the ladder operators of a system bound with the Mie-type potential in \linebreak $N$-dimensions. 

The work is arranged as follows: Section 2 gives the solution to the  $N$-dimensional Schr$\ddot{o}$dinger equation with the Mie-type potential and then obtain the energy eigenvalues and the eigenfunctions. Also, in sections 3 some expectation values and the next section deals with the ladder operators for the Mie-type potential. Finally, we give the conclusions of the work in section 5.

\pagebreak
\noindent {\bf 2.0 Schr$\ddot{o}$dinger Equation in \textsl N-dimensional Spaces.}\\

\noindent Using the \textsl N-dimensional polar cordinate with polar variable \textsl r (hyperradius) and the angular momentum variable  ${\theta_1,\theta_2,\theta_3,...,\theta_{N-2},\phi}$ (hyper angle), the Laplacian operator in the polar coordinate ${r,\theta_1,\theta_2,\theta_3,...,\theta_{N-2},\phi}$ of the $R^N$ is

$$\nabla^2_N=r^{1-N}\frac{\partial}{\partial r}\left(r^{N-1}\frac{\partial}{\partial r}\right)+ \frac{\Lambda^2_N(\Omega)}{r^2}   \eqno{(1)}$$ \\
where $\Lambda^2_N(\Omega)$   is a partial differential operator on the unit sphere $S^{N-1}$  (Laplace-Betrami operator or the grand orbital operator) define analogously to a three-dimensional angular momentum[18]  as $\Lambda^2_N(\Omega)=-\Sigma^N_{i\geq j}(\Lambda^2_{ij})$ where \linebreak $\Lambda^2_{ij}=x_i\frac{\partial}{\partial x_j}- x_j\frac{\partial}{\partial x_i}$ for all Cartesian component $x_i$ of the \textsl \linebreak \textsl N-dimensional vector $(x_1,x_2,...,x_N)$.

 The \textsl N-dimensional Schr$\ddot{o}$dinger equation has the form [1, 2, 19]:
 $$-\frac{\hbar^2}{2\mu}\nabla^2_N\Psi_{n\ell m}(r,\Omega)+V(r)\Psi_{n\ell m}(r,\Omega)=E\Psi_{n\ell m}(r,\Omega)  \eqno{(2)}$$
where $\mu$ is the reduced mass and $\hbar$ is the Plank's constant.\\

\noindent {\bf 2.1. Eigenvalues and Eigenfunctions of the Mie-Type Potential.}\\\\
The Mie-type potential, which is a diatomic potential, has been studied using methods such as the polynomial method [13], the \textsl ansatzs wave function method [15, 20] and the $1/N$ expansion method [33]. Generally, one can define the Mie-type potential as [13,34] $$V(r)=\kappa\left[\frac{a}{b-a}\left(\frac{r_e}{r}\right)^b-\frac{b}{b-a}\left(\frac{r_e}{r}\right)^a\right],  \eqno{(3)}$$where $a$ and $b$ are parameters, $\kappa$ is the interaction energy between two atoms in a molecular system at distance $r_e$. An example on this type of potentials($a=2,b=1$) is the standard Morse [21] or Kratzer-Fues [22, 23] potential of the form [15, 20, 24]
$$V(r)=-\kappa\left(\frac{2r_e}{r}-\frac{r^2_e}{r^2}\right)  \eqno{(4)}$$
 Moreover,the standard Kratzer potential is modified by adding a term to the potential.A new type of this potential is the modified Kratzer-type potential [15, 23] 
$$V(r)=-\kappa\left(\frac{r-r_e}{r}\right)^2  \eqno{(5)} $$
However,in order to give a complete solution of these two potentials, we consider a general form 
$$V(r)=-\frac{A}{r}+\frac{B}{r^2}+C  \eqno{(6)}$$
The potential appears to be more flexible due to the addition of the parameter C. With $A=\kappa r_e$,$B=\kappa r^2_e$ and $C=\kappa$,we have the modified Kratzer potential and the Kratzer-Fues potential can be obtained by setting $A=2\kappa r_e$,$B=\kappa r^2_e$ and C=0 . For brevity, inserting the potential (6) into the radial part $R_{n_r\ell}(r)$ of $\Psi_{n\ell m}(r,\Omega)$, we obtain
$$R^{\prime\prime}_{n_r\ell}(r)+\frac{N-1}{r}R^\prime_{n_r\ell}(r)- \frac{\ell(\ell+N-2)}{r^2}R_{n_r\ell}(r)+ \frac{2\mu}{\hbar^2}\left[ E+\frac{A}{r}-\frac{B}{r^2}-C\right]R_{n_r\ell}(r)=0  \eqno{(7)} $$  
where  $E$ is the energy eigenvalue, $\ell$ is the Orbital angular momentum\linebreak quantum number satisfying$$\Lambda^2_N(\Omega)Y^m_\ell(\Omega)+\ell(\ell+N-2) Y^m_\ell(\Omega)=0, \eqno{(8)}$$ 
where $Y_\ell^m (\Omega)$ are the hypersperical harmonics.

By making the substitution,
$$R_{n_r\ell}(r)=r^{-\frac{(N-1)}{2}}U_{n_r\ell}(r),    \eqno(9)$$
Equation (7) becomes 
$$U_{n_r\ell}^{\prime\prime}(r)+\left[\frac{2\mu}{\hbar^2}\left( E+\frac{A}{r}-\frac{B}{r^2}-C\right)-\frac{\ell(\ell+N-2)}{r^2}-\frac{(N-1)(N-3)}{4r^2}\right]U_{n_r\ell}(r)=0  \eqno (10)$$
We note that Equation(22)is similar to the one-dimensional Schr$\ddot{o}$dinger except for the addition of the centrifugal term $\frac{\ell(\ell+N-2)}{r^2}$ to the potential. Defining the following parameters:
$$t=\sqrt{\frac{8\mu(C-E)}{\hbar^2}}r,\hspace{5pt}\alpha=\sqrt{\frac{\mu}{2\hbar(C-E)}}A,\hspace{5pt}\nu(\nu+N-2)=\ell(\ell+N-2)+\frac{2\mu B}{\hbar^2} \eqno(11)$$
Equation(7) becomes 
$$R_{n_r\ell}^{\prime\prime}(t)+\frac{N-1}{t}R_{n_r\ell}^{\prime}(t)+\left[-\frac{1}{4}+\frac{\alpha}{t}-\frac{\nu(\nu+N-2)}{t^2}\right]R_{n_r\ell}(t)=0  \eqno(12)$$
By the behaviour of the wave function at the origin and at infinty, we define
$$R_{n_r\ell}(t)=\exp\left(-\frac{t}{2}\right)t^\nu f(t)  \eqno{(13)}$$
The use of (13) transforms (12) to the well-known assosiated Laguerre \linebreak differential equation
$$tf^{\prime\prime}(t)+\left[(2\nu+N-1)-t\right]f^\prime(t)-\left[\nu+\frac{N-1}{2}-\alpha\right]f(t)=0.  \eqno{(14)}$$ Polynomial solution to (14) can be written in terms of the hypergeometic function as follows [31]
$$f(t)=_1F_1\left(\nu+\frac{N-1}{2}-\alpha,2\nu+N-1;t\right) \eqno{(15)}$$   
For large value of $t$,this solution diverges, thus preventing normalizaion. To prevent this, we set
$$\nu+\frac{N-1}{2}-\alpha=-n_r; \hspace{5pt} n_r=0,1,2,... \eqno{(15)}$$
where $n_r$ is the hyperradial quantum number. Thus, it is clear from (11) and (15), that the energy eigenvalues can be obtain as 
$$ E_n=C-\frac{2\mu A^2}{\hbar^2[2n_r+2\nu+N-1]^2}, \hspace{0.3in} n,\ell=0,1,2,...  \eqno{(16)}$$
The energy eigenvalues obtained in (16) can be shown to be in agreement with those of the Kratzer-Fues potential by  setting $C=0$, thus, we have 
$$E_n^{KF}=-\frac{2\mu A^2}{\hbar^2[2n_r+2\nu+N-1]^2}.  \eqno{(17)}$$ Also, if $B$ and $C$ are set to be zero,(16) becomes $$E_n^{Col}=-\frac{2\mu A^2}{\hbar^2[2n_r+2\ell+N-1]^2}  \eqno{(18)}$$which is the energy values for the Coulombic-type potential [39]

\noindent Finally, using the following relation$$_1F_1(-a,b+1;x)=\frac{a!b!}{(a+b)!}L_a^b(x),  \eqno{(19)}$$ with Eqs.(11),(13) and (15),the wave function of the $N$-dimensional Schr$\ddot{o}$dinger equation for the Mie-type potential is given as
$$R_{n_r\ell}(r)=C_{n_r} r^{v-\frac{N-3}{2}} e^{-\epsilon r}L_{n_r}^{2v+1}(2\epsilon r)  \eqno{(20)}$$ where we have define
$$v=\frac{1}{2}\left[\sqrt{(2\ell+N-2)^2+\frac{8\mu B}{\hbar^2}}-1\right]  \eqno{(21)}$$ 
and 
$$\epsilon=\frac{4\mu A}{\hbar^2\left(2n_r+1+\sqrt{(2\ell+N-2)^2+\frac{8\mu B}{\hbar^2}}\right)},  \eqno{(22)}$$ and $C_{n_r}$ is the normalization constant to be determined.\\

\noindent Using the normalization condition $$\int_0^\infty|R_{n_r\ell}(r)|^2 r^{N-1}dr=1  \eqno{(23)}$$
 and the orthogonal property of the Laguerre polynomials $$\int_0^\infty z^{\eta+1}\left[L_n^\eta(z)\right]dz= \frac{(2n+\eta+1)(n+\eta)}{n!},  \eqno{(24)}$$
we obtain 
$$C_{n_r}=\sqrt{\frac{n_r!(2\epsilon)^{2v+3}}{2(n_r+v+1)(n_r+2v+1)!}}.  \eqno{(25)}$$
We can now express the normalized wave function as 
$$R_{n_r\ell}(r)=\sqrt{\frac{n_r!(2\epsilon)^{2v+3}}{2(n_r+v+1)(n_r+2v+1)!}} r^{v-\frac{N-3}{2}} e^{-\epsilon r}L_{n_r}^{2v+1}(2\epsilon r)  \eqno{(26)}$$
Therefore the complete orthonormalized energy eigenfunctions of the Mie-type potential is given as
$$\Psi_{n\ell m}(r,\theta,\varphi)=R_{n_r\ell}(r)Y_\ell^m(\Omega)  \eqno{(27)}$$
where $R_{n_r\ell}(r)$ is given in Eq.(26) and $Y_\ell^m(\Omega)$ are the hyperspherical\linebreak harmonics. Moreover, we note here that the wave function (26) reduces to those of the Coulombic-type when $B=0,C=0$ [39] and Kratzer-Fues when $C=0$ [2]. \\\\
\noindent{\bf 2.2. Degeneracy of Energy Levels.}\\\\
In this section, we give a simple presentation for determining the degeneracy of the energy levels of the Mie-type potentials in $N$-dimensions. If the potential under consideration (Eq.(6)) has no other symmetries beyond rotational invariance,the degeneracy of energy levels are therefore the multiplicities of the hyperspherical harmonics for a fixed $\nu$ (the new angular momentum defined in Eq.11).
In order to get a clear understanding of what the degenerate levels of the Mie-type potentials looks like, we write the energy equation (16) as follows
$$E_n=C-\frac{\mu A^2}{2\hbar^2\left[n+\frac{N-3}{2}\right]^2}  \eqno{(28)}$$ where $n=n_r+\nu+1$. Equation (28) is analogous to that of the Coulombic energy levels execpt for the addition of the parameter $C$.

In $N$ dimensions, the hypersperical harmonics depends on $N$-1 angular coordinates $\theta_1,\theta_2,...\theta_{N-2},\phi$ whose ranges are $0\leq\theta_j\leq\pi$ and $0\leq\phi\leq 2\pi$. Also, each hyperspherical harmonic is determined by $N$-1 integers $\nu,m_1,m_2,...m_{N-2}$. Thus by enumerating all the distinct set of $m$ value that are possible for a given $\nu$, the degeneracy is given by [39]
$$\mbox{Degeneracy}=\sum_{\nu=0}^{n-1}\sum_{-\nu}^\nu|m_1|=\sum_{\nu=0}^{n-1}(2\nu+1)=n^2  \eqno{(29)}$$ for $N=3$, and 
$$\mbox{Degeneracy}=\sum_{\nu=0}^{n-1}\sum_{m_1=0}^\nu\sum_{m_2=0}^{m_1}...\sum_{-m_{N-3}}^{m_{N-3}}|m_{N-2}|.  \eqno{(30)}$$ for $N>3$. With the help of Eqs.(29) and (30), results of the degeneracies of the Mie-type potentials in $N$-dimensions are given for are given in Table 1 for $N=3-10$ and $n=1-5$.\\\\ 
\begin{table}[t]
\caption{ \bf Degeneracies of the Mie-type Potentials in $N$ Dimensions for $N=3-10$ and $n=1-5$}
\centering
\begin{tabular}{p{.5in}c p{.5in}c p{.5in}c c c  c c }
\cline{1-6} 
$N$&$n=1$&$n=2$&$n=3$&$n=4$&$n=5$\\\cline{1-6}
3&1&4&9&16&25\\
4&1&5&14&30&55\\
5&1&6&20&50&105\\
6&1&7&27&77&182\\
7&1&8&35&112&294\\
8&1&9&44&156&450\\
9&1&10&54&210&665\\
10&1&11&65&275&935\\
\cline{1-6}
\end{tabular}

\label{tab:}
\end{table}

\noindent{\bf 3. Some Expectation Values and the Virial Theorem for the\linebreak Mie-type Potential in \textsl N-dimensions.}\\\\
In this section, we obtain the expectation values $\langle r^{-1}\rangle$ and $\langle r^{-2}\rangle$ directly using the Hellmann-Feynmann theorem (HFT)[25, 26, 27]. Although the HFT is commonly used in the calculation of intermolecular forces in molecules, however, in order to employ the HFT in calculating the expectation values, one can promote the fixed parameters which appears in the Hamiltonian to be a continuous variable (for the mathematical purpose of taking the derivative). Thus, suppose the Hamiltonian $H$ for a particular quantum system is a function of some parameters $q$, and let $E_n(q)$ and $\Psi_n(q)$ be the eigenvalues and eigenfunctions of $H(q)$ respectively, then the HFT states that$$\frac{\partial E_n(q) }{\partial q}=\langle\Psi_n(q)\arrowvert\frac{\partial H(q)}{\partial q}\arrowvert\Psi_n(q)\rangle.  \eqno{(31)}$$
The effective Hamiltonian of the hyperradial wave function is given as 
$$H=\frac{-\hbar^2}{2\mu}\frac{d^2}{dr^2}+\frac{-\hbar^2}{2\mu}\frac{(2\ell+N-1)(2\ell+N-3)}{4r^2}-\frac{A}{r}+\frac{B}{r^2}+C  \eqno{(32)}$$
To obtain $\langle r^{-1}\rangle$, we let $q=A$ such that
$$\langle\Psi_n(A)\arrowvert\frac{\partial H(A)}{\partial A}\arrowvert\Psi_n(A)\rangle=\langle r^{-1}\rangle \eqno{(33)}$$ 
and
$$\frac{\partial E_n(A) }{\partial A}=\frac{4\mu A}{\hbar^2\left[2n_r+1+\sqrt{(2\ell+N-2)^2+\frac{8\mu B}{\hbar^2}}\right]^2}  \eqno{(34)}$$
Thus by the HFT, we have
$$\langle r^{-1}\rangle=\frac{4\mu A}{\hbar^2\left[2n_r+1+\sqrt{(2\ell+N-2)^2+\frac{8\mu B}{\hbar^2}}\right]^2}  \eqno{(35)}$$
Similarly,by letting $q=B$ in Equation(28) we obtain $\langle r^{-2}\rangle$ as 
$$\langle r^{-2}\rangle=\frac{16\mu^2 A^2}{\hbar^4\sqrt{(2\ell+N-2)^2+\frac{8\mu B}{\hbar^2}}\left[2n_r+1+\sqrt{(2\ell+N-2)^2+\frac{8\mu B}{\hbar^2}}\right]^3}  \eqno{(36)}$$
We note that$q=\ell$ also yields the same result for $\langle r^{-2}\rangle$.\\
Finally,, if we let $q=\mu$, then by the HFT, we have the virial theorem of the Mie-type potential as follows:
$$-\langle H-V\rangle=(1-\beta)E_n  \eqno{(37)}$$
since $H=T+V=E_n$,we get $$ -(2-\beta)\langle T\rangle=(1-\beta)\langle V\rangle   \eqno{(38)}$$
where
$$\beta=\frac{8\mu B}{\hbar^2\sqrt{(2\ell+N-2)^2+\frac{8\mu B}{\hbar^2}}\left[2n_r+1+\sqrt{(2\ell+N-2)^2+\frac{8\mu B}{\hbar^2}}\right]}  \eqno{(39)}$$\\

\noindent {\bf 4. Ladder Operators for Mie-type Potential in \textsl N-dimensions.}\\\\
We shall now construct the creation and annihilation operator for the eigenfunctions obtained in (26) using the factorization method [28, 29, 30, 36, 37]. As shown in previous works, the ladder operators can be constructed directly from the wave function without introducing any auxiliary variable. Therefore, we intent to find differential operators $\hat{L}_\pm$  satisfying the property
$$\hat{L}_\pm R_{n_r\ell}(r)=\ell_\pm R_{n_r\pm 1,\ell}(r)  \eqno{(40)}$$\\
In other words,we wish to find the operators of the form
$$\hat{L}_\pm=f_\pm(r) \frac{d}{dr}+g_\pm (r)  \eqno{(41)}$$\\
which depend on the physical variable $r$.

To this end, we start by obtaining the derivative of the wave function (26)
$$\frac{d}{dr}R_{n_r\ell}(r)=\frac{R_{n_r\ell}(r)}{r}\left(v-\frac{N-3}{2}\right)-\epsilon R_{n_r\ell}(r)+N_{n_r}^v r^{v-\frac{N-3}{2}} e^{-\epsilon r}\frac{d}{dr}L_{n_r}^{2v+1}(2\epsilon r)  \eqno{(42)}$$
where 
$$N_{n_r}^v=\sqrt{\frac{n_r!(2\epsilon)^{2v+3}}{2(n_r+v+1)(n_r+2v+1)!}}.  \eqno{(43)}$$\\
With the use of the following relation of the asociated Laguerre function[31]
$$x\frac{d}{dx}L_n^\alpha(x)=nL_n^\alpha(x)-(n+\alpha)L_{n-1}^\alpha,  \eqno{(44)}$$
Eq.(43) becomes
$$\left[-\frac{d}{dr}-\epsilon+\frac{1}{r}\left(n_r+v-\frac{(N-3)}{2}\right)\right]R_{n\ell}(r)=\frac{(n_r+2v+1)}{r}\frac{N_{n_r}^v}{N_{n_r-1}^v}R_{n-1,\ell}(r).  \eqno{(45)}$$
Using (43),we have 
$$\left[-r\frac{d}{dr}-\epsilon r+\left(n_r+v-\frac{(N-3)}{2}\right)\right]R_{n_r\ell}(r)=\sqrt{\frac{n_r(n_r+v)(n_r+2v+1)}{(n_r+v+1)}}R_{n_r-1,\ell}(r)  \eqno{(46)}$$
Hence the annihilation operator is define as
$$\hat{L}_-=\left[-r\frac{d}{dr}-\epsilon r+\left(n_r+v-\frac{(N-3)}{2}\right)\right] \eqno{(47)}$$
with the following effect on the wave function
$$\hat{L}_- R_{n_r\ell}(r)=\ell_-R_{n_r-1,\ell}(r)  \eqno{(48)}$$ where
$$\ell_-=\sqrt{\frac{n_r(n_r+v)(n_r+2v+1)}{(n_r+v+1)}}  \eqno{(49)}$$
Similarly, the creation operator can be obtained by using the following property of the associated Laguerre polynomial [31] in Eq.(42)
$$x\frac{d}{dx}L_n^\alpha(x)=(n+1)L_{n+1}^\alpha(x)-(n+\alpha+1-x)L_n^\alpha. \eqno{(50)}$$
Thus,we have
$$\left[\frac{d}{dr}+\epsilon+\frac{1}{r}\left(n_r+v-2\epsilon r-\frac{(N-1)}{2}\right)\right]R_{n_r\ell}(r)=\frac{(n_r+1)}{r}\frac{N_{n_r}^v}{N_{n_r+1}^v}R_{n_r+1,\ell}(r).  \eqno{(51)}$$
Also,using (43), we have
$$\left[r\frac{d}{dr}-\epsilon r+\left(n_r+v-\frac{(N-1)}{2}\right)\right]R_{n_r\ell}(r)=\sqrt{\frac{(n_r+1)(n_r+v+2)(n_r+2v+2)}{(n_r+v+1)}}R_{n_r+1,\ell}(r)  \eqno{(52)}$$
whence, we can define the creator operator as
$$\hat{L}_+=\left[r\frac{d}{dr}-\epsilon r+\left(n_r+v-\frac{(N-1)}{2}\right)\right] \eqno{(53)}$$
satisfying
$$\hat{L}_+ R_{n_r\ell}(r)=\ell_+R_{n_r+1,\ell}(r)  \eqno{(54)}$$
with
$$\ell_+=\sqrt{\frac{(n_r+1)(n_r+v+2)(n_r+2v+2)}{(n_r+v+1)}}.  \eqno{(55)}$$
At this point, it is important to stress that operators (47) and (53) as they have been derived above
are not constructed as the factorizing operators of the Hamiltonian. It is well known that
the factorizing operators are not, in general, the ladders of the system [36, 37].

We now turn our attention to the algebra associated with the opperators $\hat{L}_-$ and $\hat{L}_+$. From Eqs.(47) and (53),we can compute the commutator $[\hat{L}_-,\hat{L}_+]$ as follows\\\\
$\begin{array}{lcl}
\displaystyle [\hat{L}_-,\hat{L}_+]R_{n_r\ell}(r)
&=&\hat{L}_-\{\hat{L}_+R_{n_r\ell}(r)\}-\hat{L}_+\{\hat{L}_-R_{n_r\ell}(r)\}\\\\
&=&\sqrt{\frac{(n_r+1)(n_r+v+2)(n_r+2v+2)}{(n_r+v+1)}}\{\hat{L}_-,R_{n_r+1,\ell}(r)\}\\
&&-\sqrt{\frac{n_r(n_r+v)(n_r+2v+1)}{(n_r+v+1)}}\{\hat{L}_+,R_{n_r-1,\ell}(r)\}\\\\
&=&(2n_r+2v+2)R_{n_r\ell}(r)=2\ell_0 R_{n_r\ell}(r) \hspace{.9in} (56)
\end{array}$\\
where we have introduced the eigenvalue
$$\ell_0=n_r+v+1.  \eqno{(57)}$$
We can thus define the operator
$$\hat{L}_0=\hat{n}_r+v+1   \eqno{(58)}$$
It is easy to show that the operators $\hat{L}_\pm$ and $\hat{L}_0$ satisfies the commutator relations
$$[\hat{L}_-,\hat{L}_+]=2\hat{L}_0,\hspace{.2in}[\hat{L}_0,\hat{L}_-]=-\hat{L}_-,\hspace{.2in}[\hat{L}_0,\hat{L}_+]=\hat{L}_+\eqno{(59)}$$
which correspond to the commutator relations of the dynamic group $SU(1,1)$.
Using Eq.(58), the Hamiltonian now acquires a simple form 
$$H=C-\frac{\mu A^2/2\hbar^2}{\hat{L}_0^2}.  \eqno{(60)}$$
Moreover, the following expressions can easily be obtained for the operators:
$$rR_{n_r\ell}(r)=\frac{1}{2\epsilon}[2\hat{L}_0-(\hat{L}_++\hat{L}_-)]R_{n_r\ell}(r)-\frac{N}{2\epsilon}R_{n_r\ell}(r) \eqno{(61)}$$
and 
$$r\frac{d}{dr}R_{n_r\ell}(r)=\frac{1}{2}(\hat{L}_+-\hat{L}_-)R_{n_r\ell}(r)-\frac{1}{2}R_{n_r\ell}(r)  \eqno{(62)}$$
Also,the corresponding matrix elements of these two functions can be obtained as 
$$\langle R_{n_r\ell}(r)\arrowvert r\arrowvert R_{m_r\ell}(r)\rangle=\frac{1}{2\epsilon}[(2n_r+2v+2)\delta_{m_r,n_r}-\ell_+\delta_{m_r,n_r+1}-\ell_-\delta_{m_r,n_r-1}]-\frac{N}{2\epsilon}\delta_{m,n} \eqno{(63)}$$
and
$$\langle R_{n_r\ell}(r)\arrowvert r\frac{d}{dr}\arrowvert R_{m_r\ell}(r)\rangle=\frac{1}{2}[\ell_+\delta_{m_r,n_r+1}-\ell_-\delta_{m_r,n_r-1}]-\frac{1}{2}\delta_{m_r,n_r} \eqno{(64)}$$
respectively.\\
Finally, the Casimir operator [32] can also be express as
$$\tilde{C} R_{n_r\ell}(r)=[\hat{L}_0(\hat{L}_0-1)-\hat{L}_+\hat{L}_-]R_{n_r\ell}(r)=[\hat{L}_0(\hat{L}_0+1)-\hat{L}_-\hat{L}_+]R_{n_r\ell}(r)=J(J-1)R_{n_r\ell}(r)  \eqno{(65)}$$ 
where 
$$J=v+1. \eqno{(66)}$$\\

\noindent{\bf 5. Conclusions.}\\
In this paper, the solution to the $N$-dimensional Schr$\ddot{o}$dinger equation with the Mie-type potentials was obtained. The energy eigenvalues obtained were found to be consistent with those of the 3-dimensional Mie-type potential when $N=3$, also the energy levels reduces to those of the Coulombic-type when $B=0$. The eigenfunctions were also worked out in terms of the associated Laguerre polynomials and were shown to be in agreement with those obtained in literatures as $N=3$. 

Moreover, expectation values $\langle r^{-1}\rangle$ and $\langle r^{-2}\rangle$ were calculated using the Hellmann-Feynmann theorem and the values obtained indicate the  clustered of more electrons around the Coulombic part of the potential.The viral theorem was also verified for the Mie-type potentials in $N$-dimensions and the result obtained reduces to that of the usual virial theorem $-2\langle T\rangle=\langle V\rangle$ when $B=0$, as expected for a Coulombic potential. 
Also, using the factorization method, the creation and annihilation operators were also constructed for the Mie- type potential and they were found to agree with those of the 3-dimensional Colulombic potential when $N=3$ and $B=0$ [29]; moreover, the operators were shown to obey the $SU(1,1)$ algebra. However, we stress that the operators (47) and (53) as constructed in the work are not the factorizing operators of the Hamiltonian. 

We also observed that the degeneracy of the energy levels increases with the dimension $N$. Finally, that note that with a proper choice of the parameters, the results obtained become those of the modified Kratzer and the Kratzer-Fues potentials.\\\\
{\bf Acknowledgements}: The authors wish to thank Profs. R. Sever, L.M. Nieto and O.~Rosas-Ortiz for communicating to us some of their works during the preparation of  the paper. Also, D.~Agboola wishes to acknowledge the support of Ruth Lawal, Adewale Adesola, Vincent Bankole and Tope Oderinde during the course of the work.   
 
\pagebreak

\end{document}